\documentclass[referee,a4paper,12pt,traditabstract]{swsc} 

\usepackage{color,array}
\usepackage{xcolor,url}
\usepackage{rotating}
\usepackage{amssymb}
\usepackage{tabularx,colortbl}
\usepackage[utf8]{inputenc}
\usepackage[T1]{fontenc}
\usepackage{framed}

\usepackage{graphicx}
\usepackage{txfonts}
\usepackage{subfigure}
\usepackage{epstopdf}
\usepackage[mathlines]{lineno}
\modulolinenumbers[5]
\usepackage[authoryear,round]{natbib}
\usepackage[backref]{hyperref}
\usepackage{verbatim}

\hypersetup{colorlinks=true,citecolor=cyan,urlcolor=cyan,linkcolor=blue}

\begin{document}


   \title{VOEvent for Solar and Planetary Sciences}
      
   \titlerunning{VOEvent for Solar and Planetary Sciences}

   \authorrunning{Cecconi et al.}

   \author{B. Cecconi\inst{1}
          \and
          P. Le Sidaner\inst{2}
          \and
          L. Tomasik\inst{3}
          \and
          C. Marmo\inst{4}
          \and
          M. B. Garnung\inst{5}
          \and
          J. Vaubaillon\inst{6}
          \and
          N. Andr\'e\inst{7}
          \and
          M. Gangloff\inst{7}
          }

   \institute{LESIA, Observatoire de Paris, Universit\'e PSL, CNRS, Sorbonne Universit\'e, Univ. Paris Diderot, Sorbonne Paris Cit\'e, 
              Meudon, France\\
              \email{\href{mailto:baptiste.cecconi@obspm.fr}{baptiste.cecconi@obspm.fr}}
         \and
              DIO, Observatoire de Paris, Universit\'e PSL, CNRS, Paris, France\\
              \email{\href{mailto:pierre.lesidaner@obspm.fr}{pierre.lesidaner@obspm.fr}}
	     \and 
	          Space Research Centre, Polish Academy of Sciences, Warsaw, Poland\\
	          \email{\href{mailto:tomasik@cbk.waw.pl}{tomasik@cbk.waw.pl}}
	     \and
	          GEOPS, CNRS, Universit\'e Paris-Sud, Universit\'e Paris Saclay, Orsay, France\\
	          \email{\href{mailto:chiara.marmo@u-psud.fr}{chiara.marmo@u-psud.fr}}
	     \and
	          LPC2E, Universit\'e d'Orl\'eans, CNRS,  Orl\'eans, France\\
	          \email{\href{mailto:mgarnung@cnrs-orleans.fr}{mgarnung@cnrs-orleans.fr}}
	     \and
	          IMCCE, Observatoire de Paris, Universit\'e PSL, CNRS, France\\
	          \email{\href{mailto:jeremie.vaubaillon@obspm.fr}{jeremie.vaubaillon@obspm.fr}}
         \and
         	  IRAP, CNRS, Universit\'e Paul Sabatier, Toulouse, France\\
	          \email{\href{mailto:nicolas.andre@irap.omp.eu}{nicolas.andre@irap.omp.eu}}\\
	          \email{\href{mailto:michel.gangloff@irap.omp.eu}{michel.gangloff@irap.omp.eu}}
	  }


\begin{abstract}
{With its Planetary Space Weather Service (PSWS), the Europlanet-2020-RI Research Infrastructure (EPN2020RI) project is proposing a compelling set of databases and tools to that provides Space Weather forecasting throughout the Solar System. We present here the selected event transfer system (VOEvent). We describe the user requirements, develop the way to implement event alerts, and chain those to the 1) planetary event and 2) planetary space weather predictions. The service of alerts is developed with the objective to facilitate discovery or prediction announcements within the PSWS user community in order to watch or warn against specific events. The ultimate objective is to set up dedicated amateur and/or professional observation campaigns, diffuse contextual information for science data analysis, and enable safety operations of planet-orbiting spacecraft against the risks of impacts from meteors or solar wind disturbances.}
\end{abstract}

   \maketitle
   
\section{Introduction}
The PSWS (Planetary Space Weather Service) \citep{2018P&SS..150...50A} Joint Research Activities (JRA) will set up the infrastructure necessary to transition to a full planetary space weather service within the lifetime of the project. A variety of tools (in the form of web applications, standalone software, or numerical models in various degrees of implementation) are available for tracing propagation of planetary or solar events through the Solar System and modeling the response of the planetary environment (surfaces, atmospheres, ionospheres, and magnetospheres) to those events. As these tools were usually not originally designed for planetary event prediction or space weather applications, additional development is required for these purposes. The overall objectives of PSWS will be to review, test, improve and adapt methods and tools available within the partner institutes in order to make prototype planetary event/diary and space weather services operational through PSWS Virtual Access (VA) at the end of the program. One of the goals is: \emph{To identify user requirements, develop a methodology for issuing event alerts, and link those to the planetary event and space weather predictions.} This is the scope of this paper. We first present selected science cases that demonstrate the need for the proposed system. The VOEvent infrastructure is then described, followed with the way we implement it for solar system wide space weather. 

\section{Science Cases}
The forecasting planetary and space weather events is initiated by observations. The observational events can be used as such, or used as inputs for prediction or modeling tools to predict potential subsequent effects. 

Planetary meteor impacts have been reported by several teams (including amateurs) in the last decade: First shooting star seen from Mars \citep{Selsis05}; amateur astronomer see Perseid hits on the Moon \citep{Spellman08}; Fiercest meteor shower on record to hit Mars via comet \cite{Grossman13}; Explosion on Jupiter spotted by amateur astronomers \citep{Malik12}. The events are often reported in the news or on amateur online forums. Those transient events are useful for studying the properties of the impacted region. Quick and efficient transmission of them is thus a key step. This methodology is used in astronomy with the Gamma-Ray Bursts alert system \citep{barthelmy08}. 

\begin{figure}
\centering\includegraphics[width=\linewidth]{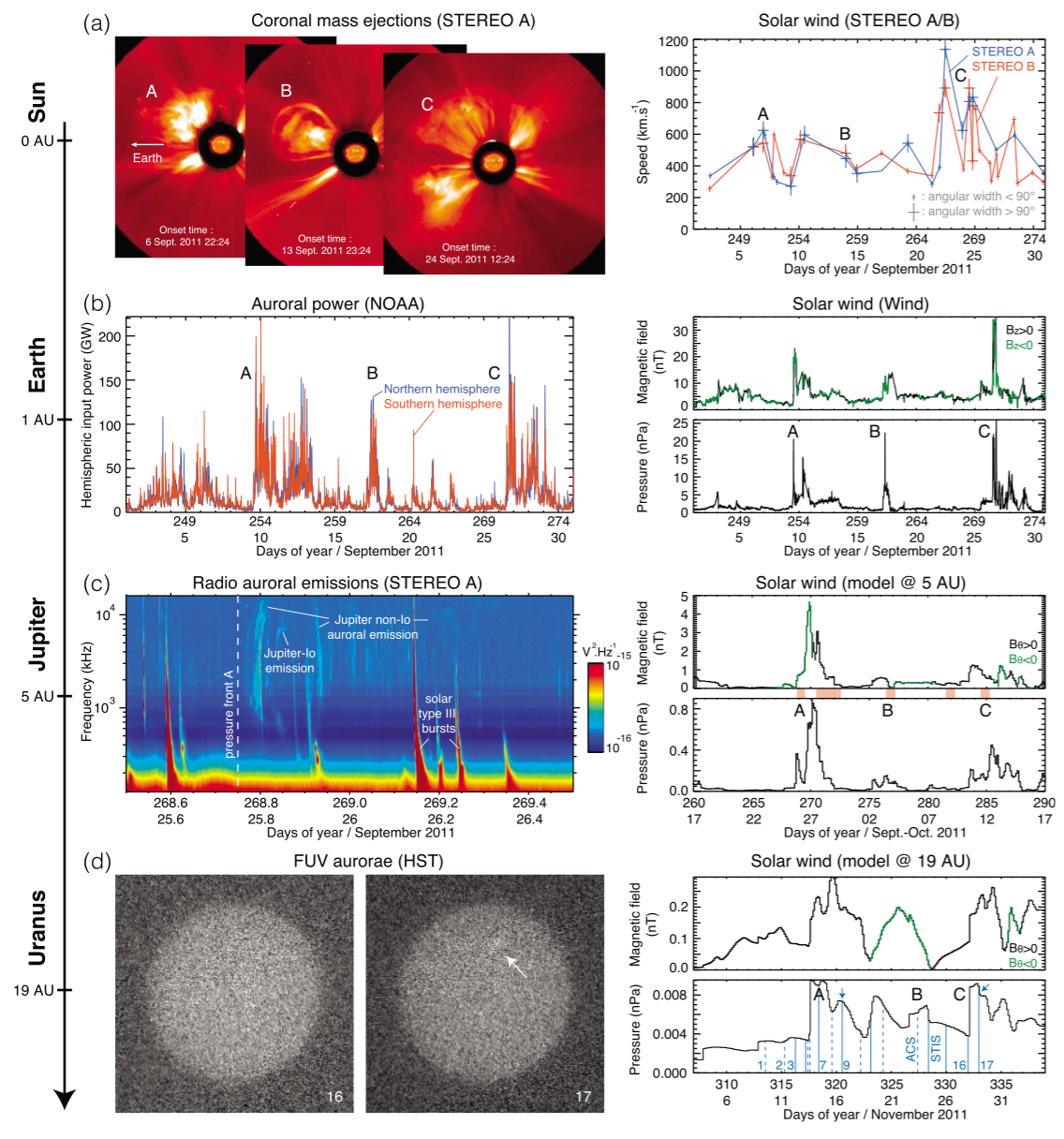}
\caption{Following an interplanetary shock throughout the solar system, from the Sun to Uranus. Figure extracted from \cite{Lamy12}}
\label{fig:ips-uranus}
\end{figure}

Several studies have been published \citep{prange04,Lamy12} presenting the observed effects of interplanetary shocks while the hit various planets throughout the solar system, from the Sun to Jupiter, Saturn or even Uranus. Figure \ref{fig:ips-uranus} shows an interplanetary shock triggered by three coronal mass ejections (CME) in September 2011. The shock has been observed at Earth a few days later (with {\it in situ} measurement on the WIND spacecraft, as well as in the auroral power monitored by NOAA). It also triggered intense decametric radio emissions at Jupiter three weeks later, that were observed by the STEREO-A/Waves instrument. It finally hit Uranus after a two months journey in the interplanetary medium, with the activation of Uranus atmospheric aurora. The planning of Uranus' aurorae has been prepared using the 1D MHD mSWiM model \citep{zieger08} developed at University of Michigan. This code also confirmed the Jovian radio detection link with the studied event. The major outcome of this study is the first observation of the faint Uranus' aurorae from Earth orbit, and this was only possible thanks to the propagation model.

\begin{figure}
\centering\includegraphics[width=\linewidth]{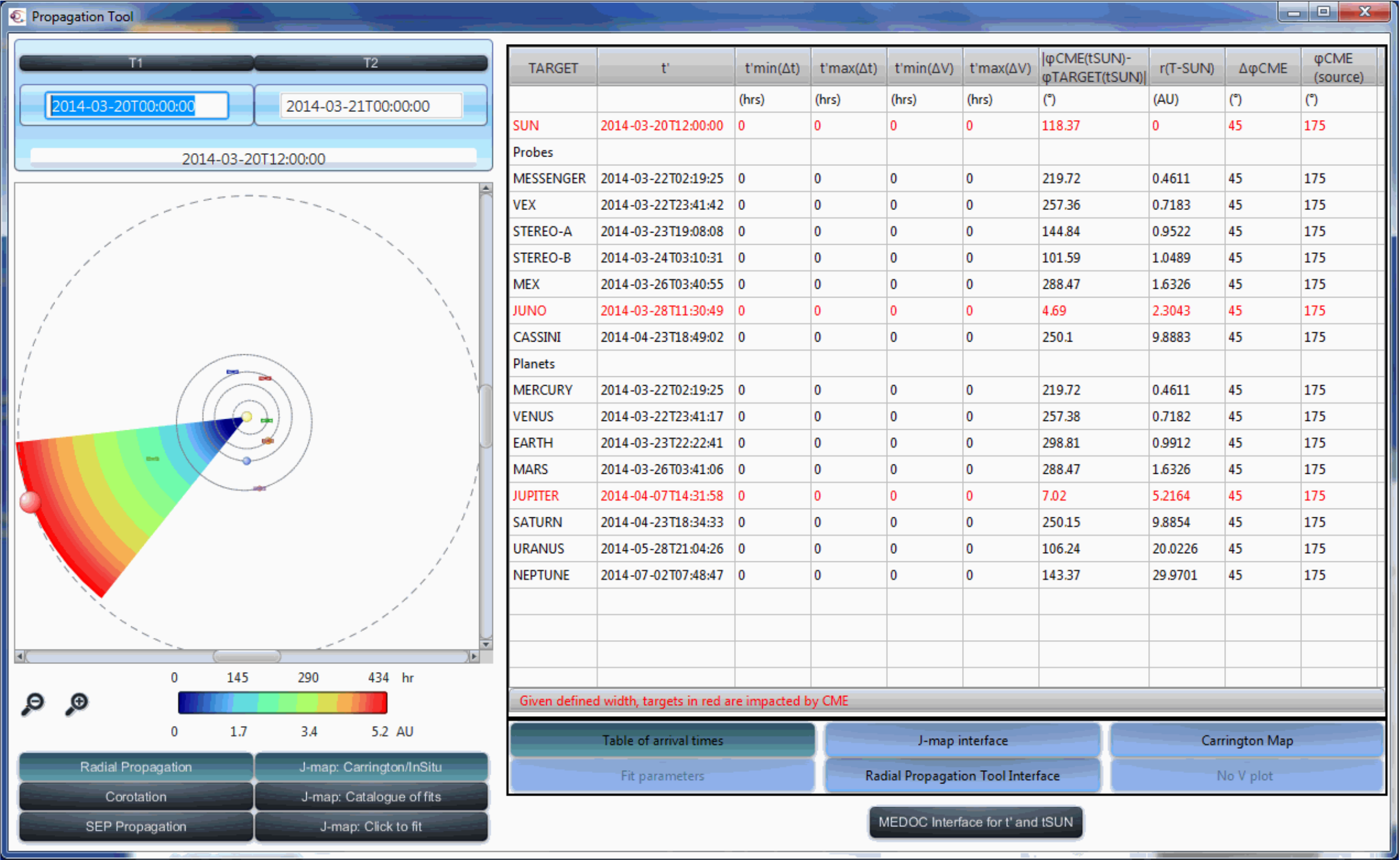}
\caption{CDPP Propagation Tool}
\label{fig:cdpp-proptool}
\end{figure}

Several heliospheric propagation models \citep[see, e.g., ][]{Tao05,zieger08} have been recently developed and provide the space physics community with time of arrivals of interplanetary shocks, or high energy particle beam at planets or spacecraft in the solar system. Online tools and repositories are providing access to heliospheric simulation runs. The Coordinated Community Modeling Center (CCMC\footnote{CCMC. \url{http://ccmc.gsfc.nasa.gov}}) is a run on demand service center with several propagation models available. Users can also use simulation runs that were previously computed. The French Plasma Physics Data Centre (CDPP\footnote{CDPP. \url{http://www.cdpp.eu}}) is providing precomputed simulation runs in his AMDA (Automated Multi Dataset Analysis) tool\footnote{CDPP-AMDA. \url{http://amda.cdpp.eu}} as time series of predicted Solar Wind parameters at the place of the various spacecraft and planets. It also proposes a Propagation Tool\footnote{CDPP Propagation Tool. \url{http://propagationtool.cdpp.eu}} that uses both simulation and observational products to derive time of arrivals of Solar Wind events at the place of the various spacecraft and planets (see Figure \ref{fig:cdpp-proptool}).

Many science teams and space missions could have taken advantage of such predictions in the recent years (Smart 1, Rosetta, MEx, MAVEN, VEx, HST, MSL, Dawn), and would take advantage of those within the next five years (Exomars, Juno, HST/JWST, Parker Solar Probe, Solar Orbiter) and on a longer term (Bepi-Colombo, JUICE). This list of missions will be used to prioritize the event catalogs, tools or models that will be implemented in the PSWS alert system.

\section{VOEvent}
VOEvent \cite{voevent} is a standardized language used to report observations of astronomical events; it was officially adopted in 2006 by the International Virtual Observatory Alliance (IVOA). Though most VOEvent messages currently issued are related to supernovae, gravitational microlensing, and gamma-ray bursts, they are intended to be general enough to describe all types of observations of astronomical events, including gravitational wave events. 

The VOEvent system is already used or planned to be used by several large-scale projects: the Gamma-Ray Coordinate System (GCN) \cite{barthelmy08}; the Large Synoptic Survey Telescope (LSST) \cite{2014htu..conf...19K}; the European Low Frequency Array (LOFAR) \cite{vanHaarlem:2013gi}; or the Solar Dynamic Observatory (SDO) \cite{Somani_AGU_2012}. That last project has scopes included in PSWS goal. In each of those projects, VOEvent is used for fast transmission of transient observations. In PSWS, we plan to use VOEvent for both observations and predictions.

Messages are written in XML, providing a structured metadata description of both the observations and the inferences derived from those observations. VOEvent messages are designed to be compact and quickly transmittable over the internet via the VOEvent Transfert Protocol (VTP). The version of the VOEvent standard is 2.0, at the time of writing.

\begin{figure}
\centering\includegraphics[width=0.7\linewidth]{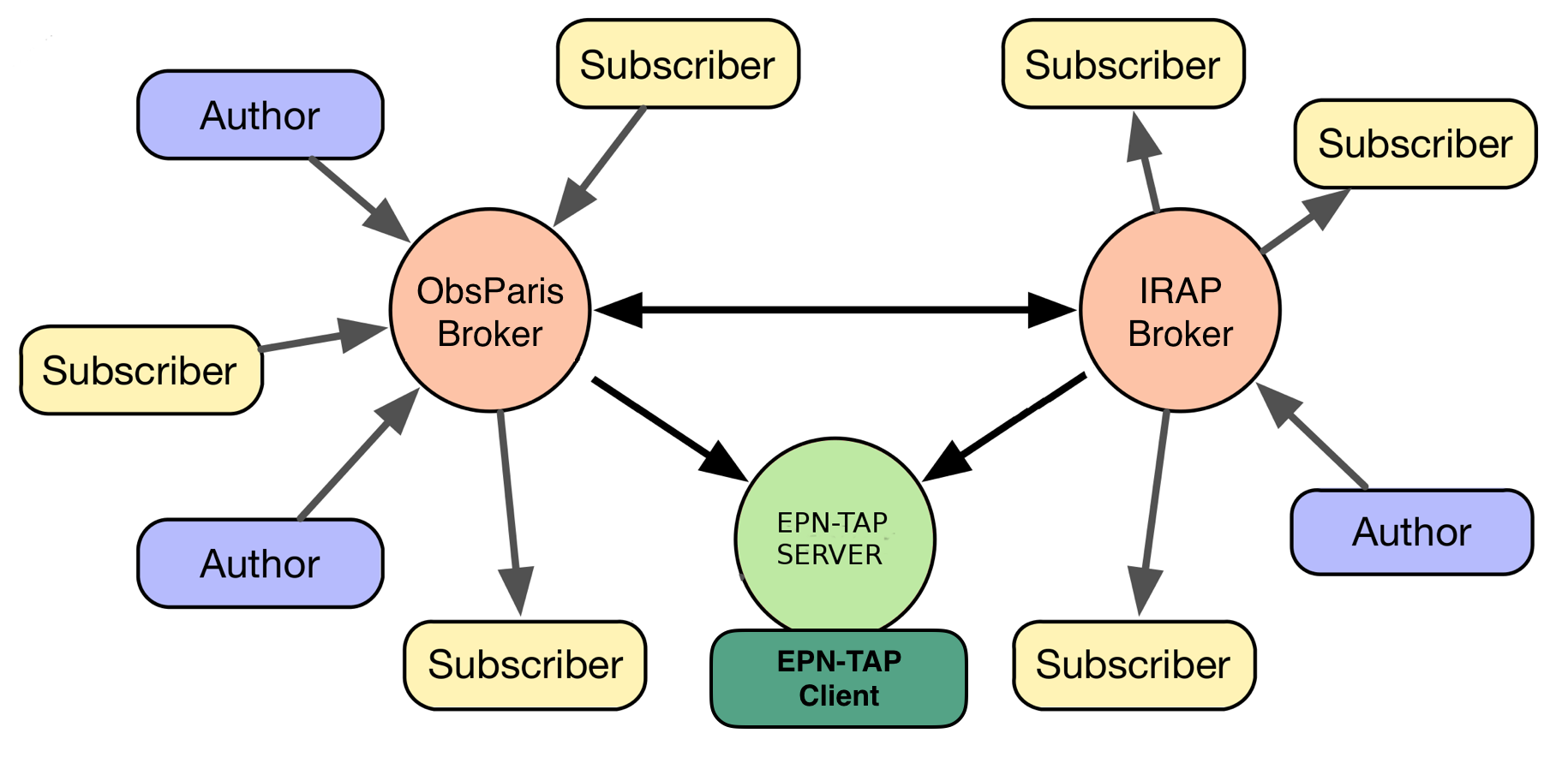}
\caption{PSWS VOEvent Network Architecture (adapted from VOEvent documentation available at \url{https://voevent.readthedocs.io/en/latest/receiving.html})}
\label{fig:architecture}
\end{figure}

As shown on Figure \ref{fig:architecture}, there are three types of nodes: \emph{Author}, \emph{Broker} and \emph{Subscriber}.  The \emph{Authors} are issuing VOEvents. The \emph{Brokers} are dispatching the VOEvents received from \emph{Authors} to \emph{Subscribers}. \emph{Subscribers} are receiving VOEvents from \emph{Brokers}. The large-scale network is composed of a series of \emph{Brokers} that are also \emph{Subscribers} of other \emph{Brokers}. The \emph{Authors} must assign a unique IVOA identifier \cite{ivoid} to each issued VOEvent. The \emph{Subscribers} are configured to only take into account VOEvents with new identifiers. In order to update an event (e.g., update the predicted time of an event, after improved processing), a new VOEvent must be issued as an update of a previous VOEvent with reference to the previous VOEvent identifier. This system ensures consistency and avoids conflicting messages. 

A VOEvent message contains the following tags:
{\tt <who>} Describing who is responsible (the author and the publisher) for the information contained in the message;
{\tt <what>} the data (such as source flux) associated with the observations of the event;
{\tt <wherewhen>} description of the time and place where the event was recorded. This draws from the Space-time Coordinate (STC) recommendation to the IVOA;
{\tt <how>} a description of the instrumental setup on where the data were obtained;
{\tt <why>} inferences about the nature of the event;
{\tt <description>} a description of the event content.
A well-formed VOEvent message must validate against the VOEvent XML schema. A valid message may omit most of the informational tags listed above, but since the creation of VOEvent messages is done automatically, most opt to transmit the fullest content available.

\section{Adjusting VOEvent for Heliophysics and Planetary Sciences}
The main difference between astronomical and Solar System events is the reference frames in use. Sky coordinates (Right Ascension and Declination) are the obvious choice for astronomy, and they have been implemented in the VOEvent model. The {\tt <wherewhen>} part of VOEvent is currently not fully usable for the variety of reference frames required for Solar System events. 

The current assessment of Solar System requirements for the evolution of the {\tt <wherewhen>} section revealed the following new needs: (a) Capability to specify the only target name as a location. It can be a planet, a satellite, a comet, a moon, a spacecraft, a rover, etc. Standard names should be used here, such as IAU names for natural bodies. (b) Capability to specify the location using coordinate frame on a body, or in a reference frame related to a object in the Solar System. (c) Capability to specify a time range, with a start and end time. The list of reference frames (including its time scale, its coordinate system and its spatial origin) will be compiled by the VESPA and PSWS teams. 

When dealing with predictions, the usage of the {\tt <how>} and {\tt <why>} section is proposed as follows: the {\tt <how>} part covers tool or model used to derive the prediction; while {\tt <why>} concerns the original event or observation used as input of how to derive prediction, as well as the possible threshold met on predicted parameters for issuing the event.

Modifying the VOEvent data model (and XML schema implementation) to handle target names and target types as well as any Solar System coordinate systems is a task that requires coordination with several working groups of the IVOA. The IVOA Solar System Interest Group (SSIG) is in charge on including Solar System needs into IVOA models and will propose evolution of the standards. Before this is fully implemented, Solar System events specific parameters are handled inside the {\tt  <What>} part of VOEvent using the {\tt <group>} capability of the VOEvent XML schema. This is also the current implementation option selected by high energy astrophysics and gravitational wave events, which do not fit perfectly in the VOEvent model either. We propose here a specific implementation solution to define Solar System parameters into VOEvents using the current Schema version 2.0.

\section{Europlanet PSWS and VESPA infrastructure for VOEvent}

The PSWS team has build a redundant broker using Paris Astronomical Data Centre (PADC) and IRAP/CDPP facilities.  The broker is implemented using the \emph{Comet} open-source server available from \url{https://github.com/jdswinbank/Comet} \cite{Swinbank:2014ft}. The author nodes can also be implemented with \emph{Comet} to submit VOEvents. On the client side, a lighter listener using the \emph{pygcn} package (available from \url{https://github.com/lpsinger/pygcn}) is used to listen to a specific broker. 

As VOEvent is a message sent once and with no persistent trace, we use the VESPA infrastructure to store event metadata together with a link to the original VOEvent XML file. Each event will be referred to with its {\tt ivorn} identifier, which must be unique. Each VOEvent author source have a dedicated EPNcore metadata catalogue table reachable with TAP (Table Access Protocol), following VESPA (Virtual European Solar and Planetary Access) recommendation \cite{vespa}.

This allows to make data mining in previously posted VOEvents (observations or predictions), e.g., look for meteor shower at Mars during the full 2020 year. The corresponding ADQL \cite{adql} query would be:
\begin{verbatim}
SELECT * FROM meteor_shower.epn_core WHERE shower_year = '2000' AND 
target_name = 'Mars' 
\end{verbatim}
and should sent to the TAP service access point (see Table \ref{tab-services}). The EPNcore metadata catalogue table contains the standard EPNcore keywords as well as extra parameters related to the VOEvent specificities. In the example shown above, the {\tt target\_name} keyword is part of the EPNcore standard, whereas {\tt shower\_year} is specific to the {\tt meteor\_shower} PSWS service.

\section{Example Implementations}

We present here four example implementations of various readiness levels showcasing the variety of topics and needs for the planetary space weather events. It is noticeable, that two of the examples are concerning Earth related observations or predictions, but the teams are also considering this infrastructure for broadcasting events.

\begin{table}
\begin{tabular}{lll}
Service Title&TAP server Access URL&schema name\\
\hline
Meteor Shower on Planets & \url{http://voparis-tap-voevent.obspm.fr/tap} & {\tt meteor\_shower}\\
Space-Weather Forecast & \url{http://vespa.cbk.waw.pl/tap} & {\tt rwcalerts}\\
\end{tabular}
\caption{Examples of prototype VOEvent EPN-TAP services}
\label{tab-services}
\end{table}

\subsection{Meteor Showers on Planets}
The Meteor Shower Prediction code is using the IMCCE celestial mechanics models to predict the occurrence of meteor showers at various planets. The ephemeris are produced by simulating the ejection of meteoroid from the sunlit hemisphere of cometary nuclei, typically from 0 to 3 AU, followed by the propagation of orbits of meteoroids in 
the Solar System, taking into account the gravitation of the Sun, the 8 planets, Pluto and the Moon, 
as well as the radiation pressure and the Poynting-Robertson drag. The showers are predicted when a 
 planet enters a large enough number of meteoroids, at a distance less than typically 0.01 AU \cite{2005A&A...439..751V, 2005A&A...439..761V}. 

The Meteor Shower alerts contains information on the shower: its source (a named comet), its target (a planet) and various shower parameters, including the direction of arrival (in ICRS RA Dec coordinates) of the meteoroid stream as seen from the center of the target body. The prediction also provides the user with temporal information as well as a confidence index parameter \cite{2017P&SS..143...78V}.

Appendix \ref{appendix-xml-voevent-meteor} shows an example of a VOEvent derived from the simulation. The {\tt <why>} section is filled with references to the Unified Astronomy Thesaurus \cite{2014ASPC..485..461A}. An alternative implementation is also presented in Appendix \ref{appendix-xml-voevent-meteor} , using the {\tt <WhereWhen>} entity. 

\subsection{Space-Weather Forecast}
The International Space Environment Services (ISES) is a space weather service organization currently comprised of 16 Regional Warning Centers, which main objective is to provide services to the scientific and user communities within their own regions. These services usually consist of forecasts or warnings of disturbances to the solar terrestrial environment.

Polish RWC Warsaw meets this task through the web based communication channels like http and email protocols based on traditional codes and human readable text messages, created by the Heliogeophysical Prediction Service Laboratory. The data processing system (Helgeo2) is responsible for gathering data, performing simulation and preparation of first/provisional forecast that will be almost instantly manually reviewed by an Helgeo2 operator and published to the users. 

Users focused on system automatization (scientific and commercial, such as single LOFAR stations, ionosonde, low orbit satellite and power grid operators) requires data inputs easy to process by the IT system like well formatted XML structure in VOEvent, therefore the daily alerts with forecast of solar and geomagnetic activity have been translated. 

The RWC Warsaw VOEvent daily alerts contain Solar activity and geomagnetic activity forecasts included in separated  {\tt <what>} groups. The Solar activity forecast provides a predicted level of activity for the Sun with a scale of 5 flags (very low, low, moderate, high, very high), and a probability of Solar flares in percent for C, M, X and P flare classes. The geomagnetic activity forecast provides a predicted KP index. The {\tt <why>} section contains the rationale and observations that led to the forecast. In this section, the ESPAS \cite{Belehaki:2016bq} ontology is used for the concept definitions.

The database of RWC Warsaw VOEvent is available through the web page \url{http://helgeossa.cbk.waw.pl/DATA/rwc/alerts/} and by EPN-TAP service (see Table \ref{tab-services}). An example of VOEvent file is displayed in Appendix \ref{appendix-xml-voevent-rwc}.

\subsection{Prediction of Solar Wind Parameters}
\begin{figure}
\includegraphics[width=\linewidth]{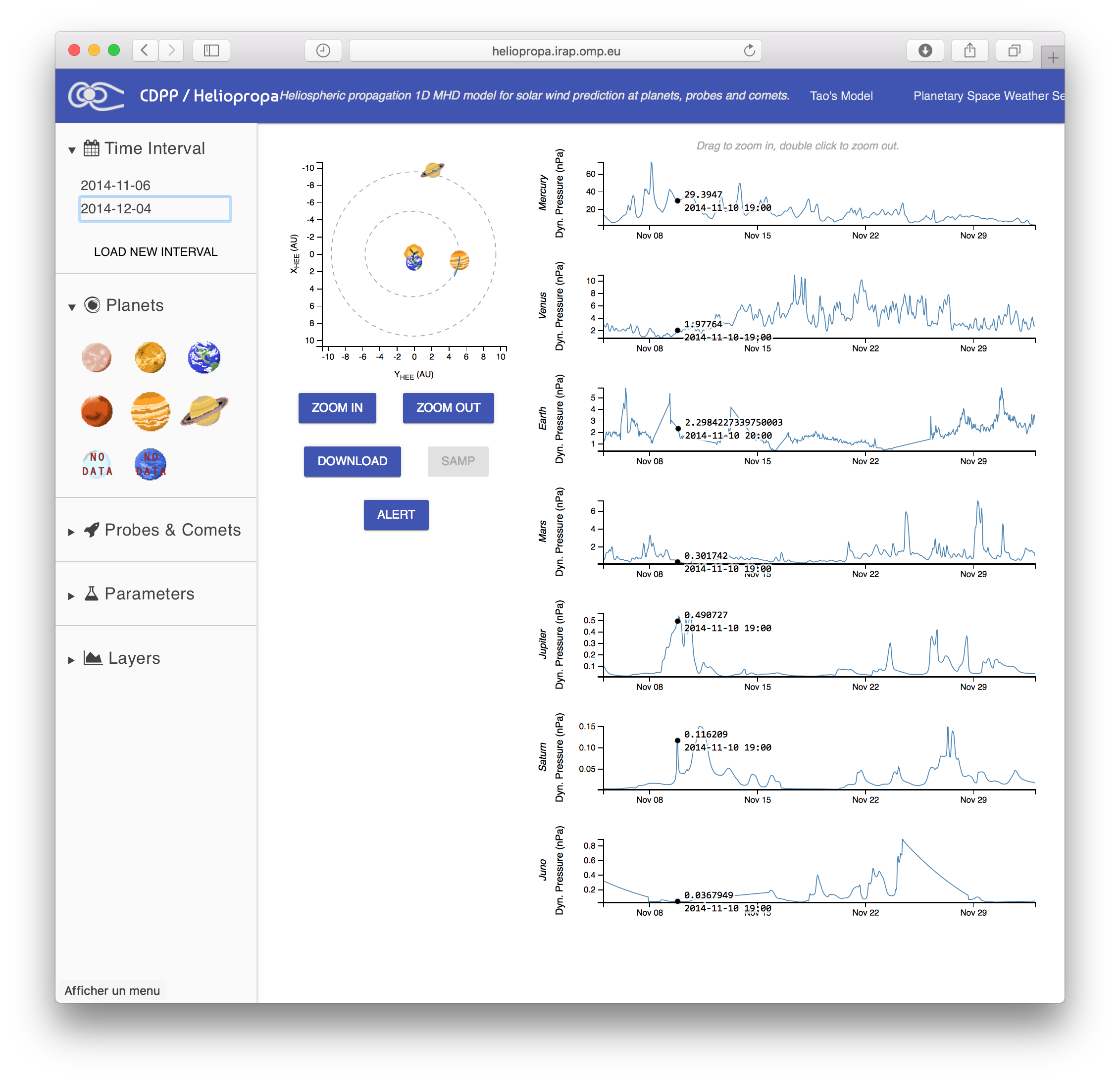}
\caption{Screenshot of the HelioPropa tool developed by CDPP.}
\label{fig-heliopropa}
\end{figure}

This prediction tool is developed by the IRAP/CDPP team. The model is based on 1D-MHD propagation code developed by \cite{doi:10.1029/2004JA010959}. The code in using the in-situ measurements of Solar Wind parameters in the Earth vicinity, as available from the OMNI website (\url{http://nssdc.gsfc.nasa.gov/omniweb/}). The predicted Solar Wind parameter profiles are available through the HelioPropa tool developed by the CDPP (\url{http://heliopropa.irap.omp.eu}, see Fig., \ref{fig-heliopropa}), and events can be issued when a certain threshold is passed on predefined quantities. 

For the prototype available at the time of writing, users have to register and setup the threshold value and the tested parameter in order to receive alerts. The alerts are sent by email in a VOTable file \cite{2013ivoa.spec.0920O}. The service is currently available here: \url{http://alerts-psws.irap.omp.eu/tao}.

Appendix \ref{appendix-xml-voevent-tao} shows a VOEvent based on predefined thresholds selected by the CDPP science team, containing the same information as the current VOTable output. This VOEvent example makes use of the {\tt <table>} element to define list of values.

\subsection{Optical Detection of Meteors and Transient Luminous Events}
Optical counterparts of atmospheric transient events are subject of observation and investigation of a growing community. Professional as well as amateur observation networks are established in Europe and abroad to collect data important not only for the scientific investigation of some physical processes but also to assess their possible impacts on Earth's environment. We propose here an implementation of the VOEvent standard to the domain of meteors and Transient Luminous Events (TLEs). A well established standard for real-time alert on those domains will facilitate coordination between networks and simplify the extraction of critical information. This would result in improved collaboration between the amateur community and agencies, e.g., the ESA Fireball Database\footnote{\url{http://neo.ssa.esa.int/search-for-fireballs}}, or the CNES TARANIS (Tool for the Analysis of RAdiations from lightNIngs and Sprites) mission \cite{2007AdSpR..40.1268B}. 

Transient Luminous Events (TLEs) \cite{Pasko2012}  are large-scale optical events occurring in the upper-atmosphere from the top of thunderclouds up to the ionosphere. They are sometimes accompanied by terrestrial gamma-ray flashes (TGFs). TLEs may have important effects in local, regional, and global scales of the atmosphere, but many features of TLEs are not fully understood yet. TARANIS is a CNES satellite project dedicated to the study of impulsive transfers of energy between the Earth atmosphere and the space environment. The TARANIS microsatellite will fly over thousands of TLEs and TGFs for at least two years. Its scientific instruments will be capable of detecting these events and recording their luminous and radiative signatures, as well as the electromagnetic perturbations they set off in Earth's upper atmosphere. Coupling TLEs observation to the already existent meteor detection networks, will allow the observation of TLEs over unprecedented space and time scales \cite{2017AGUFMAE23A2469G}, strongly increasing the probability of joint detection and hence the scientific return of space missions such as TARANIS or the ESA ASIM (Atmosphere-Space Interactions Monitor) instrument onboard the ISS (International Space Station) \cite{2009AIPC.1118....8N}.

Several camera networks already exist in Europe and around the world, aiming to detect and triangulate shooting stars, compute the trajectory of the possible meteorite and constrain the orbital properties of the meteoroid. Professional and amateur networks (see among others \cite{1998M&PS...33...49O,  2013pimo.conf..125J, 2015EPSC...10..800C, 2016pimo.conf...76G}) working together will allow Europe to be completely independent in obtaining awareness about Earth space environment and existing risks connected to atmospheric reentries. European Space Situational Awareness national programs would benefit from having a common and standard framework for sharing information on meteor and fireball detections, and their contribution to the ESA Fireball Information System would become more efficient. A Virtual Meteor Observatory initiative is already existing in a European context \cite{2008EM&P..102..247K, 2010JIMO...38...10B}, ending up in the adoption of an XML-based communication format. Its connection to VOEvent standard will guarantee its sustainability and interoperability in a larger and well documented context.

Appendices \ref{appendix-xml-voevent-meteor-freeture} and \ref{appendix-xml-voevent-tle-freeture} show preliminary VOEvent messages for a meteor and a TLE detection, planned to be implemented in the meteor and TLE detection FreeTure \cite{2014pim4.conf...39A} software. 

\section{Discussion and Perspectives}

VOEvent is used as a standard in Astronomy for fast advertising new event on the sky. In this study, we show that Solar System sciences can take advantage of this infrastructure without any change, even if the VOEvent data model has not been designed for Solar and planetary sciences. The adjustments of the VOEvent model for Solar System use case is minimal and should be harmless for current implementations, as it just adds terms and concept, without changing those current in use. The Solar System Interest Group of IVOA is working on implementing these modifications with the Time Domain Interest Group, which is managing the VOEvent standard. 

Several projets have expressed interest in studying and possibly using VOEvent for their alert system. We mention here the LOFAR4SW design study.
The LOFAR4SW project (funded by H2020 EC framework programme) is currently conducting a design study for a future LOFAR measurement and processing pipeline dedicated to Space Weather. This study includes hardware developments for a Space Weather dedicated measurement backend, as well as all aspects of the processing down to the data distribution. The team is studying the VOEvent and EPN-TAP infrastructure proposed by the PSWS and VESPA teams for the future LOFAR4SW alert system. The current implementation of RWC is proving the feasibility in this domain. 

As a final word, we plan to reach the general public with broadcasting part of the VOEvent messages through Twitter, using the content of the {\tt <Description>} field. This would ensure a wider audience, especially with the amateur community.

\acknowledgements{The authors are supported by the Europlanet 2020 Research Infrastructure, which has received funding from the European Union's Horizon 2020 research and innovation programme under grant agreement No 654208.}

\bibliography{voevent}

\appendix

\section{VOEvent examples}

\subsection{Example for a Meteor Shower alert at Mars}
\label{appendix-xml-voevent-meteor}

VOEvent example for a Meteor Shower alert at Mars produced by a trail of comet 2006 AR3. The maximum of event is predicted to occur on 2018-09-08T18:25:00.

{\tiny 
\verbatiminput{./voevent-meteor.xml}
}

Possible implementation of the {\tt <WhereWhen>} entity in the Meteor Shower prediction example. This implementation is currently not valid because the \emph{astroCoordSystemID} required here is not in the enumerated list of allowed values in the VOEvent model

{\tiny 
\begin{verbatim}
<?xml version="1.0" encoding="UTF-8"?>
<voe:VOEvent ivorn="ivo://vodpc.obspm/imcce/meteorshower/event#MAR_AR3_2018_1102_0::v1.0"
    role="prediction" version="2.0"
    xmlns:xsi="http://www.w3.org/2001/XMLSchema-instance"
    xmlns:voe="http://www.ivoa.net/xml/VOEvent/v2.0"
    xsi:schemaLocation="http://www.ivoa.net/xml/VOEvent/v2.0 http://www.ivoa.net/xml/VOEvent/VOEvent-v2.0.xsd">
    [...]
    <WhereWhen>
        <ObsDataLocation>
            <ObservatoryLocation/>
            <ObservationLocation>
                <AstroCoordSystem id="UTC-ICRS-MARS_C"/>
                <AstroCoords>
                    <Time>
                        <TimeInstant>
                            <ISOTime>2018-09-08T18:25:00Z</ISOTime>
                        </TimeInstant>
                    </Time>
                    <Position2D unit="deg">
                        <Value2>
                            <C1>345.422</C1>
                            <C2>-16.112</C2>
                        </Value2>
                        <Error2Radius>0.204</Error2Radius>
                    </Position2D>
                </AstroCoords>
            </ObservationLocation>
        </ObsDataLocation>
    </WhereWhen>
    [...]
</voe:VOEvent>
\end{verbatim}
}

\subsection{Example for RWC daily forecast}
\label{appendix-xml-voevent-rwc}

VOEvent example for RWC daily forecast on September 7th 2017.

{\tiny 
\begin{verbatim}
<?xml version="1.0" encoding="utf-8"?>
<voe:VOEvent xmlns:voe="http://www.ivoa.net/xml/VOEvent/v2.0"
   xmlns:xsi="http://www.w3.org/2001/XMLSchema-instance"
   ivorn="ivo://src.pas/ssa/voe/wa2voe/170906wa/0" role="prediction" version="2.0"
   xsi:schemaLocation="http://www.ivoa.net/xml/VOEvent/v2.0 http://www.ivoa.net/xml/VOEvent/VOEvent-v2.0.xsd">
   <Who>
      <AuthorIVORN>ivo://src.pas/organization</AuthorIVORN>
      <Date>2017-09-06T09:52:31</Date>
   </Who>
   <What>
      <Description>Forecast of Solar-geophysical activity and propagation conditions</Description>
      <Reference uri="http://rwc.cbk.waw.pl/rwc/170906wa.dre"/>
      <Group name="Solar activity forecast">
         <Description>1 day forecasts</Description>
         <Param name="target_name" value="Sun" ucd="meta.id" utype="epn:target_name" dataType="string"/>
         <Param name="time_min" value="2017-09-07T00:00:00" ucd="time.start" dataType="string" utype="epn:time_min"/>
         <Param name="time_max" value="2017-09-08T00:00:00" ucd="time.end" dataType="string" utype="epn:time_max"/>
         <Param name="activity" ucd="meta.code.class" unit="" value="high(4)" dataType="string"/>
         <Param name="flares class C probability" ucd="stat.probability" unit="%" value="100"/>
         <Param name="flares class M probability" ucd="stat.probability" unit="%" value="100"/>
         <Param name="flares class X probability" ucd="stat.probability" unit="%" value="50"/>
         <Param name="flares class p probability" ucd="stat.probability" unit="%" value="95"/>
      </Group>
      <Group name="Magnetic activity forecast">
         <Description>1 day forecasts</Description>
         <Param name="target_name" value="Earth" ucd="meta.id" utype="epn:target_name" dataType="string"/>
         <Param name="time_min" value="2017-09-07T00:00:00" ucd="time.start" dataType="string" utype="epn:time_min"/>
         <Param name="time_max" value="2017-09-08T00:00:00" ucd="time.end" dataType="string" utype="epn:time_max"/>
         <Param name="Kp" ucd="meta.code.class;phys.magfield" unit="" value="5" dataType="string"/>
      </Group>
   </What>
   <WhereWhen/>
   <How>
      <Description>Solar-geophysical prediction service RWC Warsaw - Regional Warning Centre of ISES</Description>
      <Reference uri="http://rwc.cbk.waw.pl"/>
   </How>
   <Why importance="1.0" expires="2017-09-07T00:00:00">
      <Concept>Solar and Geomagnetic Observations</Concept>
      <Inference probability="1.0">
         <Name>Solar Activity</Name>
         <Concept>http://ontology.espas-fp7.eu/phenomenon/SolarActivity</Concept>
         <Description>
            Activity    : moderate(3)
            Flares      : 5 class m, m2.3/1n 17.37-17.51, m3.8 06.33-06.43,
            : m3.2 04.33-05.07, m4.2 01.03-01.11ut, c5.4 10.13-10.25ut,
            : 2 class 1, 16 class 0
         </Description>
      </Inference>
      <Inference probability="1.0">
         <Name>Geomagnetic activity</Name>
         <Concept>http://ontology.espas-fp7.eu/phenomenon/GeomagneticActivity</Concept>
         <Description>
            activity    : low k=2
            storm       : -
         </Description>
      </Inference>
   </Why>
</voe:VOEvent>
\end{verbatim}
}

\subsection{Example for Solar Wind event at Jupiter}
\label{appendix-xml-voevent-tao}
VOEvent example for Solar Wind dynamic pressure pulses at Jupiter, as modeled by the Tao 1D MHD propagation code \cite{Tao:2005dp}, for October 2018

{\tiny 
\begin{verbatim}
<?xml version="1.0"?>
<voe:VOEvent ivorn="ivo://psws.irap/VOEvent/Tao_Jupiter_2018-10-02T17_34_45::v1.0" role="prediction"
   version="2.0" xmlns:xsi="http://www.w3.org/2001/XMLSchema-instance"
   xmlns:voe="http://www.ivoa.net/xml/VOEvent/v2.0"
   xsi:schemaLocation="http://www.ivoa.net/xml/VOEvent/v2.0 http://www.ivoa.net/xml/VOEvent/VOEvent-v2.0.xsd">
   <Who>
      <AuthorIVORN>ivo://psws</AuthorIVORN>
      <Author>
         <contactEmail>Michel.Gangloff@irap.omp.eu</contactEmail>
         <contactName>Michel Gangloff</contactName>
      </Author>
      <Date>2018-10-02T17:34:45</Date>
   </Who>
   <What>
      <Description>Time intervals in which dynamic pressure is greater than 0.08 nPa at Jupiter</Description>
      <Param name="event_type" value="Solar Wind dynamic pressure pulse" ucd="meta.id"/>
      <Group name="event">
         <Param name="tested_parameter" value="Dynamic Pressure" ucd="meta.id" dataType="string"/>
         <Param name="threshold" value="0.08" ucd="phys.pressure" unit="nPa" dataType="string"/>
      </Group>
      <Group name="target">
         <Param name="target_name" value="Jupiter" ucd="meta.id" utype="epn:target_name"/>
         <Param name="target_class" value="planet" ucd="meta.id" utype="epn:target_class"/>
         <Param name="target_region" value="Magnetosphere" ucd="meta.id" utype="epn:target_region"/>
      </Group>
      <Table>
         <Description>Time intervals found by the propagation code</Description>
         <Field dataType="string" name="Start Time" ucd="time.begin" utype="epn:time_min">
            <Description>time tag for beginning of interval</Description>
         </Field>
         <Field dataType="string" name="Stop Time" ucd="time.end" utype="epn:time_max">
            <Description>time tag for end of interval</Description>
         </Field>
         <Data>
            <TR>
               <TD>2018-10-03T10:00:00.000</TD>
               <TD>2018-10-04T01:00:00.000</TD>
            </TR>
            <TR>
               <TD>2018-10-04T07:00:00.000</TD>
               <TD>2018-10-04T11:00:00.000</TD>
            </TR>
            <TR>
               <TD>2018-10-08T01:00:00.000</TD>
               <TD>2018-10-08T07:00:00.000</TD>
            </TR>
         </Data>
      </Table>
   </What>
   <WhereWhen/>
   <How>
      <Description>This prediction has been computed by the Heliospheric propagation 1D MHD model for solar wind 
         prediction at planets, probes and comets.
      </Description>
      <Reference uri="https://onlinelibrary.wiley.com/doi/10.1029/2004JA010959/abstract"/>
   </How>
   <Why>
      <Inference probability="1.0">
         <Concept>http://astrothesaurus.org/uat/1534</Concept>
         <Name>Solar Wind measured at 1 AU</Name>
         <Reference uri="https://omniweb.gsfc.nasa.gov"/>
      </Inference>
      <Inference>
         <Concept>http://astrothesaurus.org/uat/1966</Concept>
         <Name>MHD simulation of propagated Solar Wind at Jupiter</Name>
         <Reference uri="https://onlinelibrary.wiley.com/doi/10.1029/2004JA010959/abstract"/>
      </Inference>
   </Why>
   <Description>3 Solar Pressure pulses predicted at Jupiter between 2018-10-03 and 2018-10-09</Description>
</voe:VOEvent>
\end{verbatim}
}

\subsection{Example for a meteor detection}
\label{appendix-xml-voevent-meteor-freeture}
VOEvent example for a meteor detection using the meteor and TLE detection FreeTure \cite{2014pim4.conf...39A} software.

{\tiny 
\begin{verbatim}
<?xml version="1.0" encoding="UTF-8"?>
<voe:VOEvent xmlns:xsi="http://www.w3.org/2001/XMLSchema-instance"
    xmlns:voe="http://www.ivoa.net/xml/VOEvent/v2.0"
    xsi:schemaLocation="http://www.ivoa.net/xml/VOEvent/v2.0
    http://www.ivoa.net/xml/VOEvent/VOEvent-v2.0.xsd"
    version="2.0" role="test" ivorn="ivo://io.freeture.myobs/meteor_events#123">
    <Who>
        <AuthorIVORN>ivo://freeture.io.myobs</AuthorIVORN>
        <Date>2018-03-20T00:03:21</Date>
        <Author>
            <title>FreeTure Meteor VOEvent</title>
            <contactName>Pinco Pallino</contactName>
        </Author>
    </Who>
    <What>
        <Group name="source_flux">
            <Param dataType="float" name="ADU_peak_flux" ucd="phot.count" unit="count"
                value="0.0015">
                <Description>Instrumental Peak Flux</Description>
            </Param>
            <Param dataType="float" name="cal_peak_mag" ucd="phot.mag" unit="mag" value="10.8">
                <Description>Calibrated Peak Magnitude</Description>
            </Param>
        </Group>
        <Table name="light_curve">
            <Description>Detection properties</Description>
            <Param name="zero_point" dataType="float" ucd="phot.calib;arith.zp" unit="mag" value="23.6"/>
            <Param name="color_term" dataType="float" ucd="phot.color;arith.diff" unit="mag" value="0.037"/>
            <Param name="optical_center" dataType="float" ucd="pos.cartesian;pos.wcs.crpix" unit="pixel" value="648,440"/>
            <Param name="center_coords" dataType="float" ucd="pos.eq;pos.wcs.crval" unit="deg" value="18.34,44.52"/>
            <Param name="cd_matrix" dataType="float" ucd="pos.angResolution;pos.wcs.cdmatrix" unit="deg/pixel" value="1.17e-1,0,0,1.17e-1"/>
            <Param name="event_duration" dataType="float" ucd="time.duration" unit="s" value="0.04"/>

            <Field name="x" ucd="pos.cartesian.x;meta.main" dataType="float" unit="pixel"/>
            <Field name="y" ucd="pos.cartesian.y;meta.main" dataType="float" unit="pixel"/>
            <Field name="RA" ucd="pos.eq.ra;meta.main" dataType="float" unit="deg"/>
            <Field name="Dec" ucd="pos.eq.dec;meta.main" dataType="float" unit="deg"/>
            <Field name="alt" ucd="pos.az.alt;meta.main" dataType="float" unit="deg"/>
            <Field name="azi" ucd="pos.az.azi;meta.main" dataType="float" unit="deg"/>
            <Field name="Flux" ucd="phot.count;meta.main" dataType="float" unit="count"/>
            <Field name="Mag" ucd="phot.mag;meta.main" dataType="float" unit="mag"/>
            <Field name="Time" ucd="time.epoch;meta.main" dataType="string"/>
            <Field name="FrameId" ucd="meta.id;meta.main" dataType="int"/>
            <Data>
                <TR>
                    <TD>780.68</TD>
                    <TD>410.27</TD>
                    <TD>18.34</TD>
                    <TD>44.16</TD>
                    <TD>35.45</TD>
                    <TD>223.46</TD>
                    <TD>225.78</TD>
                    <TD>23.15</TD>
                    <TD>2018-03-19T23:55:21.2702</TD>
                    <TD>31</TD>
                </TR>
                <TR>
                    <TD>781.68</TD>
                    <TD>410.27</TD>
                    <TD>18.34</TD>
                    <TD>44.16</TD>
                    <TD>35.45</TD>
                    <TD>223.46</TD>
                    <TD>225.78</TD>
                    <TD>23.15</TD>
                    <TD>2018-03-19T23:55:21.2802</TD>
                    <TD>32</TD>
                </TR>
                <TR>
                    <TD>782.68</TD>
                    <TD>410.27</TD>
                    <TD>18.34</TD>
                    <TD>44.16</TD>
                    <TD>35.45</TD>
                    <TD>223.46</TD>
                    <TD>225.78</TD>
                    <TD>23.15</TD>
                    <TD>2018-03-19T23:55:21.2902</TD>
                    <TD>33</TD>
                </TR>
                <TR>
                    <TD>783.68</TD>
                    <TD>410.27</TD>
                    <TD>18.34</TD>
                    <TD>44.16</TD>
                    <TD>35.45</TD>
                    <TD>223.46</TD>
                    <TD>225.78</TD>
                    <TD>23.15</TD>
                    <TD>2018-03-19T23:55:21.3002</TD>
                    <TD>34</TD>
                </TR>
                <TR>
                    <TD>784.68</TD>
                    <TD>410.27</TD>
                    <TD>18.34</TD>
                    <TD>44.16</TD>
                    <TD>35.45</TD>
                    <TD>223.46</TD>
                    <TD>225.78</TD>
                    <TD>23.15</TD>
                    <TD>2018-03-19T23:55:21.3102</TD>
                    <TD>35</TD>
                </TR>
            </Data>
        </Table>
    </What>
    <WhereWhen>
        <ObsDataLocation>
            <ObservatoryLocation id="myobs">
                <AstroCoordSystem id="UTC-GEOD-TOPO"/>
                <AstroCoords coord_system_id="UTC-GEOD-TOPO">
                    <Position3D unit="deg deg m">
                        <Value3>
                            <C1>248.4056</C1>
                            <C2>31.9586</C2>
                            <C3>2158</C3>
                        </Value3>
                    </Position3D>
                </AstroCoords>
            </ObservatoryLocation>
            <ObservationLocation>
                <AstroCoordSystem id="UTC-ICRS-TOPO"/>
                <AstroCoords coord_system_id="UTC-ICRS-TOPO">
                    <Time unit="s">
                        <TimeInstant>
                            <ISOTime>2018-03-19T23:55:21.3657</ISOTime>
                        </TimeInstant>
                    </Time>
                </AstroCoords>
            </ObservationLocation>
        </ObsDataLocation>
    </WhereWhen>
    <How>
        <Description>FreeTure Temporal Detection Method</Description>
    </How>
    <Why>
        <Concept>process.variation.meteor;atmosphere.entry</Concept>
        <Description>Possible meteor event</Description>
    </Why>
</voe:VOEvent>
\end{verbatim}
}

\subsection{Example for a Trans-Luminous Event detection}
\label{appendix-xml-voevent-tle-freeture}
VOEvent example for a meteor detection using the meteor and TLE detection FreeTure \cite{2014pim4.conf...39A} software.

{\tiny 
\begin{verbatim}
<?xml version="1.0" encoding="UTF-8"?>
<voe:VOEvent xmlns:xsi="http://www.w3.org/2001/XMLSchema-instance"
    xmlns:voe="http://www.ivoa.net/xml/VOEvent/v2.0"
    xsi:schemaLocation="http://www.ivoa.net/xml/VOEvent/v2.0 http://www.ivoa.net/xml/VOEvent/VOEvent-v2.0.xsd"
    version="2.0" role="test" ivorn="ivo://freeture.io.myobs/tle_events#123">

    <Who>
        <AuthorIVORN>ivo://freeture.io.myobs</AuthorIVORN>
        <Date>2018-03-20T00:03:21</Date>
        <Author>
            <title>FreeTure TLE VOEvent</title>
            <contactName>Matthieu Garnung</contactName>
            <contactEmail>matthieu.garnung@cnrs-orleans.fr</contactEmail>
            <contributor>S.Celestin, sebastien.celestin@cnrs-orleans.fr</contributor>
        </Author>
    </Who>

    <What>
        <Group name="event parameters">
            <Param dataType="float" name="Median position x" ucd="pos.cartesian.x;meta.main"
                unit="pixel" value="307.56"/>
            <Param dataType="float" name="Median position y" ucd="pos.cartesian.y;meta.main"
                unit="pixel" value="258.97"/>
        </Group>
        <Group name="instrument">
            <Param dataType="string" name="camera_model" ucd="meta.id" value="DMK"/>
            <Param dataType="string" name="network" ucd="meta.ref" value="mycameranetwork"/>
            <Param dataType="string" name="sensor_model" ucd="instr.det;meta.id" value="ICX"/>
            <Param dataType="float" name="pixel_size" ucd="instr.pixel" value="3.5e-6" unit="m"/>
            <Param dataType="int" name="frame_size" ucd="instr.param" value="1280,960" unit="px"/>
            <Param dataType="int" name="bit_depth" ucd="instr.param" value="12" unit="bits"/>
            <Param dataType="string" name="color_flag" ucd="instr.para;meta.code" value="False"/>
        </Group>
    </What>

    <WhereWhen>
        <ObsDataLocation>
            <ObservatoryLocation id="myobs">
                <AstroCoordSystem id="UTC-GEOD-TOPO"/>
                <AstroCoords coord_system_id="UTC-GEOD-TOPO">
                    <Position3D unit="deg deg m">
                        <Value3>
                            <C1>248.4056</C1>
                            <C2>31.9586</C2>
                            <C3>2158</C3>
                        </Value3>
                    </Position3D>
                </AstroCoords>
            </ObservatoryLocation>
            <ObservationLocation>
                <AstroCoordSystem id="UTC-ICRS-TOPO"/>
                <AstroCoords coord_system_id="UTC-ICRS-TOPO">
                    <Time unit="s">
                        <TimeInstant>
                            <ISOTime>2018-03-19T23:55:21.3657</ISOTime>
                        </TimeInstant>
                    </Time>
                </AstroCoords>
            </ObservationLocation>
        </ObsDataLocation>
    </WhereWhen>
    <How>
        <Description>DMK camera from mycameranetwork, using ICX sensor.</Description>
    </How>
    <Why>
        <Description>Possible transient luminous event</Description>
    </Why>
</voe:VOEvent>
\end{verbatim}
}

\end{document}